# Crossover from normal (*N*) Ohmic subdivision to superconducting (*S*) equipartition of current in parallel conductors at the *N-S* transition: Theory

N. Kumar

Raman Research Institute, Bangalore 560080, India

**Abstract**

The recently observed[1] equipartition of current in parallel at and below the Normal-Superconducting (*N-S*) transition can be understood in terms of a Landau-Ginzburg order-parameter phenomenology. This complements the explanation proposed earlier[1] based on the flux-flow resistance providing a nonlinear negative current feedback towards equipartition when the transition is approached from above. The present treatment also unifies the usual *textbook* inductive subdivision expected much below $T_c$, and the equipartition as $T_c$ is approached from below. The question of metastability is also briefly discussed.

FOLLOWING the recent work[1], consider two, or more than two pieces of a wire of a superconducting material (*e.g.,* Nb-Ti), of unequal lengths in general, connected in parallel and fed from a constant current source. Let the combination, assumed to be in the normal state initially, be now cooled to and below the *N-S* transition temperature $T_c$. It was experimentally observed[1] that the usually expected normal ohmic subdivision of the current crosses over to an unexpected superconducting equipartition at the transition ($T_c$), and remains so divided below $T_c$. The crossover was found to be reversible. A physical explanation for this self-organization (namely, the equipartition without any fine tuning) was provided in terms of a nonlinear negative current feedback that drove the current division towards equalization. This involved a nonlinear flux-flow resistance (*R*) due to flux de-pinning by the transport current (*J*) giving $\partial R/\partial J > 0$. In the present work, we propose a complementary treatment in which the *N-S* transition is approached from the superconducting side (from below $T_c$). It is based on a Landau-Ginzburg order-parameter phenomenology in the presence of non-zero supercurrent. Our analytical treatment gives the equipartition as $T_c$ is approached starting from an arbitrary initial current division much below $T_c$. It also resolves some other puzzles associated with the phenomenon, *e.g.,* the inductive equilibrium subdivision of the supercurrent much below $T_c$, and also the question of metastability.

Label the two conductors connected in parallel as *A* and *B*. Let $\ell_A$ and $\ell_B$ be the respective lengths, unequal in general. For simplicity, the cross-sectional areas are taken to be equal (as in the experiment[1]), and assumed unity. Further, the transverse linear dimensions are assumed $<<\xi(T)$ (coherence length) and $\lambda(T)$ (the penetration length), which is certainly the case close to $T_c$. This ensures homogeneity of the superconducting order parameter over the entire conducting system. For a total supercurrent $J (=J_A + J_B)$, the dominant *G-L* free energy functional[2]

$$F = F_A + F_B$$

with
$$F_A = \ell_A \left( f_N + \alpha |\psi_A|^2 + \frac{\beta}{2} |\psi_A|^4 + \frac{|\psi_A|^2}{2} mV_A^2 \right), \quad (1)$$

and similarly for $F_B$. Here the superfluid velocity $V_{A,B}$ is related to the current (density) $J_{A,B}$ as

$$J_{A,B} = 2e |\psi_{A,B}|^2 V_{A,B} \quad (2)$$

with the constraint $J_A + J_B = J$.

Now, we must minimize $F$ variationally with respect $\psi_A$, $\psi_B$, $J_A$ and $J_B$ with the above current constraint.

Consider first the simple case of an arbitrary current division $J_A$, $J_B$ well below $T_c$. The order parameters are then essentially independent of the currents. It is sufficient then to consider current variations. From $\partial F / \partial J_A = 0 = \partial F / \partial J_B$, with $J_A + J_B = J$, we at once obtain the current ratio

$$J_A / J_B = \ell_B / \ell_A \propto \frac{\text{self-induction of } B}{\text{self-induction of } A} \quad (3)$$

as noted in textbooks.[3] Thus, indeed the supercurrent divides inductively at equilibrium well below the transition temperature. (Recall, that for equal cross-sections, self-inductance is proportional to the length. Here, we have ignored the mutual inductance).

Next, we consider the case of real interest here when we approach the *N-S* transition from below. Then the order parameters $\psi_A$ *and* $\psi_B$ dominate variationally, giving

$$\alpha(T) + \beta |\psi_A|^2 + \frac{1}{2} mV_A^2 = 0, \quad (4)$$

and similarly for $\psi_B$. Recall now from the Landau theory of second-order phase transitions that $\alpha(T) = \alpha_0 (1 - T/T_{co})$ as $T \to T_{co}$ from below, with $T_{co}$ the thermodynamic critical temperature in the absence of currents.

Introduce dimensionless quantities, $\chi_{A,B} = |\psi_{A,B}| / |\psi_0|$, $j_{A,B} = J_{A,B} / J_0$ with $|\psi_0|^2 = (-\alpha/\beta)$ and $J_0 = \left( \frac{2\sqrt{2} e \alpha_0^{3/2}}{\beta m^{1/2}} \right)$. Eliminating the velocity $V_{A,B}$ in favour of the current $J_{A,B}$, Eq. (4) gives

$$j_{A,B} = (1 - T/T_{co})^{3/2} \chi_{A,B}^2 (1 - \chi_{A,B}^2)^{1/2}. \quad (5)$$

Now, $\chi^2 (1 - \chi^2)^{1/2} \leq 2/(3\sqrt{3})$, and maximizes at $\chi = (2/3)^{1/2}$. With this, Eq. (5) defines a (normalized) critical current

$$j_c = (2/3\sqrt{3})(1-t)^{3/2} \quad (6)$$

such that $j_{A,B} > j_c$ drives the conductors A,B normal. Here $t=T/T_{co}$ is the dimensionless reduced temperature. This naturally leads to a universal temperature-current phase diagram as shown in Figure 1, and gives equipartition at *N-S* transition as follows.

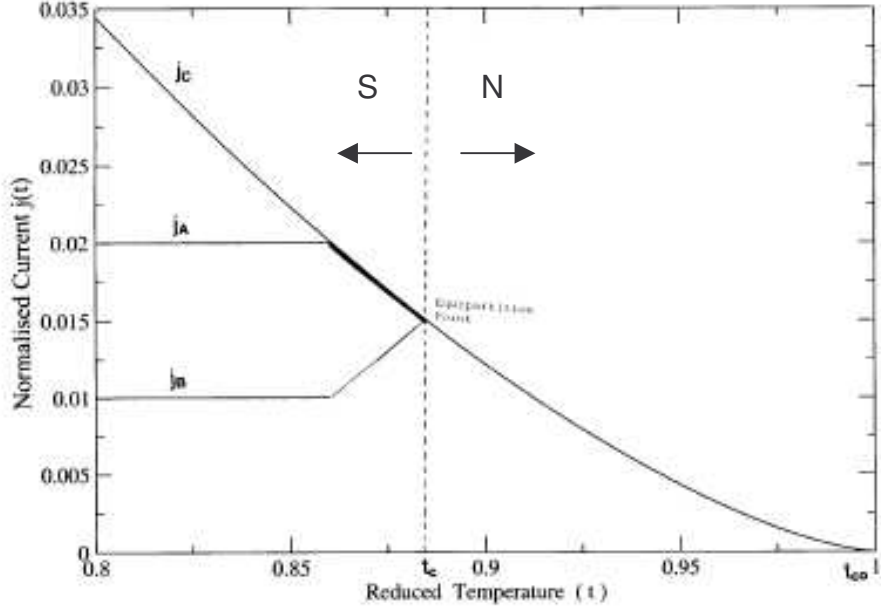

**Figure 1.** Current-Temperature phase diagram showing approach to equipartition $j_A=j_B=j/2$ at the reduced critical temperature $t_c$. Here $j_A$ and $j_B$ are the normalized currents in the parallel conductors A and B, and $j_C(t)$ is the normalized critical current. Here $t_{co}$ is the thermodynamic transition temperature at zero current.

Start well below the *N-S* transition with an arbitrary supercurrent (*j*) division, $j = j_A + j_B$, between the two conductors A and B connected in parallel with $j_A>j_B$, say. In the plot (see Fig. 1) we have taken $j_A= 0.02$ and $j_B = 0.01$ with the total current $j = 0.03$. (Recall that in the present treatment with the conductors having equal cross-sectional areas, the current is the same as the current density). The thermodynamic critical current (density) $j_C(t)$, however, is a material specific property, and its temperature dependence is given by Eq.(6). Now, as the temperature is raised towards criticality ($t_{co}$), the critical current decreases monotonically towards zero. Inasmuch as $j_A>j_B$, we must have the line $j_A$ meet the curve $j_C(t)$ first as shown in Fig. 1. Thereafter, $j_A$ must necessarily remain equal to and follow the critical current $j_C(t)$. Thus, $j_A$ and $j_C(t)$ decrease together with increasing temperature along the thick line as shown in the figure. The constraint $j_A + j_B = j$ (constant) then requires $j_B$ to rise and eventually meet the thick line at the "*equipartition point*" at $t = t_{co}$ as marked in the figure. Finally, for $t>t_{co}$, of course, both the conductors A and B are driven normal, and the current division necessarily jumps to the Ohmic subdivision in the normal state. Thus, the supercurrent equipartition at the transition point ($t_{co}$) is a universal consequence of the *N-S* criticality.

Some general remarks are in order now. The above Ginzburg-Landau phenomenological treatment should hold for both, the type-I as well as the type-II superconductors having sharp *N*-S transition. For systems with broad transition, *e.g.,* granular superconductors, a more detailed treatment will be required that takes into account the flux de-pinning and flux flow in particular. This should be particularly interesting as it may be possible then to track the fine structure of approach to equipartition at the transition. Further, it should be noted that we have assumed the transverse dimensions of the conductors to be much smaller than the coherence length and the penetration depth. This condition should certainly hold close to the transition point. This would imply no flux trapping in the loop formed by the two conductors in parallel. Much below the transition, however, the flux trapping is expected for thick conductors leading to metastable states, *i.e.,* the current division will depend on whether the conductors are pre-cooled before impressing the current from an external source, or the sample is post-cooled after the current has been impressed.

In conclusion, we have given an order parameter phenomenological treatment for the equipartition of supercurrent between two conductors connected in parallel as the *N-S* transition is approached from below. The treatment can be readily generalized to include any number of parallel conductors. We should also add that similar equipartition should clearly hold for superfluid *HeII* flowing through capillaries in parallel.

ACKNOWLEDGEMENT. The author would like to express his thanks to Prof. S.V. Bhat for many early discussions on the phenomenon of equipartition, and to Mr. Dibyendu Roy for a critical reading of the manuscript. He would also like to thank the Department of Atomic Energy (DAE), India, for a Homi Bhabha Chair.

1. Sarangi, S., Chockalingam, S.P., Mavinkurve, R.G., Bhat, S.V. and Kumar, N., Equipartition of current in parallel conductors on cooling through the superconducting transition, *J. Phys.*: *Condens. Matter*, 2006, **18**, L143-L147; *also, see* http://arXiv.cond-mat/0510099.
2. de Gennes, P.G., Superconductivity of metals and alloys, W.A. Benjamin Inc., New York, 1966.
3. Poole, C.P. Jr., Farach, H.A., and Creswick, R.J., Superconductivity (New York, Academic, 1995, p.456.